\documentclass[a4paper,aps,pre,reprint,superscriptaddress]{revtex4-1}

\usepackage[latin1]{inputenc}    
\usepackage{amsmath}           
\usepackage{amsfonts}          
\usepackage{amssymb}           
\usepackage{lmodern,dsfont}	
\usepackage{graphicx}          
\usepackage[T1]{fontenc}       

\begin{document}
\title{Pulsed chaos synchronization in networks with adaptive couplings}
\author{Marco Winkler}
\author{Sebastian Butsch}
\author{Wolfgang Kinzel}
	\affiliation{Institute for Theoretical Physics, University of Würzburg, Am Hubland, 97074 Würzburg, Germany}
\date{\today}
\begin{abstract}
Networks of chaotic units with \textit{static} couplings can synchronize to a common chaotic trajectory. The effect of \textit{dynamic adaptive} couplings on the cooperative behavior of chaotic networks is investigated. The couplings adjust to the activities of its two units by two competing mechanisms: An exponential decrease of the coupling strength is compensated by an increase due to de-synchronized activity. This mechanism prevents the network from reaching a steady state. Numerical simulations of a coupled map lattice show chaotic trajectories of de-synchronized units interrupted by pulses of mutually synchronized clusters. These pulses occur on all scales, sometimes extending to the entire network. Clusters of synchronized units can be triggered by a small group of synchronized units.
\end{abstract}

\maketitle


\section{Introduction}

Networks of interacting nonlinear units show interesting cooperative properties, for example synchronization, dynamic clusters and scale-free activity. Therefore, a lot of recent research has been invested in understanding the relation between microscopic mechanisms and macroscopic behavior in nonlinear networks \cite{Boccaletti2006175,ArenasReview}. These results are of fundamental interest in nonlinear dynamics with a wide range of applications from coupled lasers to neural networks.

In particular, synchronization of irregular spiking in neural networks is considered to be important for processing information in the brain \cite{Fries2005474,Womelsdorf_Schoffelen_Oostenveld_Singer_Desimone_Engel_Fries_2007,buzsaki2006rhythms,Dehaene2011200}. Obviously, the brain does not relax to a state of complete synchrony. In fact, the resting state of brain tissue exhibits activity on all sizes with a maximal range of response to external stimuli \cite{Beggs2003,Plenz_2010}. Synchronization, criticality and irregular activity of a neural network are generated by adaptive and competing synapses.

Networks of nonlinear units with adaptive couplings have been investigated before \cite{gross2009adaptive,kurths06,itoKaneko01,Ravoori09,Zhigulin,Ott09,Gorochowski10,aoki09,seliger02}. Simple models of excitable units with adaptive rewiring rules relax to a state of criticality \cite{Bornholdt2000,Levina09,Meisel09}. Already linear networks with adaptive couplings, following a negative Hebbian rule, show unexpected cooperative behavior; many linear modes are competing with each other \cite{Magnasco2009}. If the Hebbian rule is limited by a global restriction, a neural network develops a modular structure with a scale-free distribution of coupling strengths \cite{Gutierrez2011}.

Synchronization and irregular activity are not mutually excluded. Networks of identical nonlinear units can synchronize to a common chaotic trajectory \cite{PecoraCarroll,synchronizationBook,ArenasReview}. For networks with static couplings, the stability of chaos synchronization is related to the spectral properties of the underlying graph topology \cite{PecoraCaroll98,synchronizationBook}. Here we extend this work to networks with dynamic couplings.

In this paper we introduce a simple model - a network of coupled maps - which shows synchronization, chaos and scale-free activity. These cooperative properties are generated by a local adaptation rule which is governed by two competing mechanisms: A slow component prevents complete synchronization and a fast one which increases correlation between interacting chaotic units. Chaos synchronization is an unstable solution of our model equations.

We introduce adaptive couplings which change according to the current degree of synchronization among the units and thus influence the stability of the synchronization manifold. If the two units connected by a coupling are synchronized, the coupling strength decreases, whereas for large de-synchronization the coupling becomes enhanced. We find that the adaptive network is still chaotic. It shows pulses of chaos synchronization on all scales; the distributions of sizes and durations of the pulses heavy tailed. The strengths of the couplings are distributed around the critical value of the corresponding static network.


\section{Model}

For the sake of simplicity, initially we investigate completely connected networks of iterated maps. Each unit is described by a variable $x^i_t \in [0,1],\, i=1,..,N$ which develops in discrete time steps $t$  according to the equation 
\begin{equation} \label{eq:dynamics}
	x_{t}^{i} = \left(1-\sum_{j\neq i}\frac{\epsilon_{t-1}^{ij}}{N-1} \right)\,f\left(x_{t-1}^{i}\right)+\sum_{j\neq i}\frac{\epsilon_{t-1}^{ij} }{N-1}\,f\left(x_{t-1}^{j}\right),
\end{equation}
where the couplings $\epsilon_{t}^{ij} \in \left[0,1\right]$ are time dependent. In this contribution we use the skew tent map
\begin{equation}
	f(x) = 
		\begin{cases} \frac{x}{\alpha}, & x < \alpha\\
                  \frac{1-x}{1-\alpha}, & x \geq \alpha
    \end{cases}
\end{equation}
with $\alpha= \frac{3}{5}$ in order to model the chaotic behavior of the units, but we found analogous results for the logistic map \footnote{$f\left(x\right) = r \, x \, \left(1-x\right)$  ,  $r=4$} and the Bernoulli map \footnote{$f\left(x\right) = \left(a \, x \right) \text{mod} \, 1$  ,  $a=\frac{5}{3}$}.

The synchronization manifold $x^i_t=s_t$ is a solution of equations \eqref{eq:dynamics},
\begin{equation} \label{eq:sm}
	s_t = f\left(s_{t-1}\right).
\end{equation}
Hence, $s_t$ has a chaotic trajectory with the Lyapunov exponent  $\lambda = \frac{3}{5} \, \text{ln} \frac{5}{3} + \frac{2}{5} \, \text{ln} \frac{5}{2} \approx 0.673$.

The stability of this synchronization manifold can be calculated using the master stability function and the eigenvalues of the coupling matrix \cite{PecoraCaroll98}. For \textit{static} couplings, $\epsilon^{ij}_t=\epsilon$, the synchronization is stable if  the coupling is large enough. One finds \cite{synchronizationBook}
\begin{equation} \label{eq:couplingThreshold}
	\frac{1-e^{-\lambda}}{1+\frac{1}{N-1}} < \epsilon < \frac{1+e^{-\lambda}}{1+\frac{1}{N-1}}.
\end{equation}
In the thermodynamic limit, $N \to \infty$, this becomes
\begin{equation} \label{eq:border}
	0.490 < \epsilon < 1.510
\end{equation}
with the relevant synchronization border $\epsilon_c \approx 0.490$. Yet, already for $N=100$ the critical coupling is quite similar, namely $\epsilon_c = 0.485$.

However, here we consider \textit{dynamic} couplings. The network \eqref{eq:dynamics} has  $N(N-1)$ couplings with time dependent strengths $\epsilon^{ij}_t$; each coupling changes according to the activity of the two units which it connects. We consider two competing mechanisms: On the one hand, the coupling is slowly driven to zero. On the other hand, de-synchronized units increase the strength of their mutual coupling. The dynamics of the couplings is defined as
\begin{equation} \label{eq:adaptation}
	\begin{split}
		\epsilon_{t}^{ij} & = \left(\mu+\sigma\,\left(\Delta_{t-1}^{ij}\right)^2\right)\,\epsilon_{t-1}^{ij}\\
		\Delta^{ij}_t &= \left|x_t^{i} - x_t^{j}\right|.
	\end{split}
\end{equation}
In order to keep the couplings bounded, we reset each to $\epsilon^{ij}_t=1$ if it leaves the unit interval. In the synchronization manifold, where $\Delta^{ij}=0$, the coupling decays to zero exponentially in time with the relaxation time $\tau= -\frac{1}{\text{ln}\mu} \approx \frac{1}{1-\mu}$ for $\mu \lesssim 1$. If the couplings are small enough, any perturbation of the synchronized trajectory drives the system apart from chaos synchronization. But then the second mechanism tries to increase the couplings again to enforce synchronization.

The competition of these two effects leads to a complex cooperative behavior of the whole network.

\begin{figure}
	\centering
	\includegraphics[width = 0.46\textwidth]{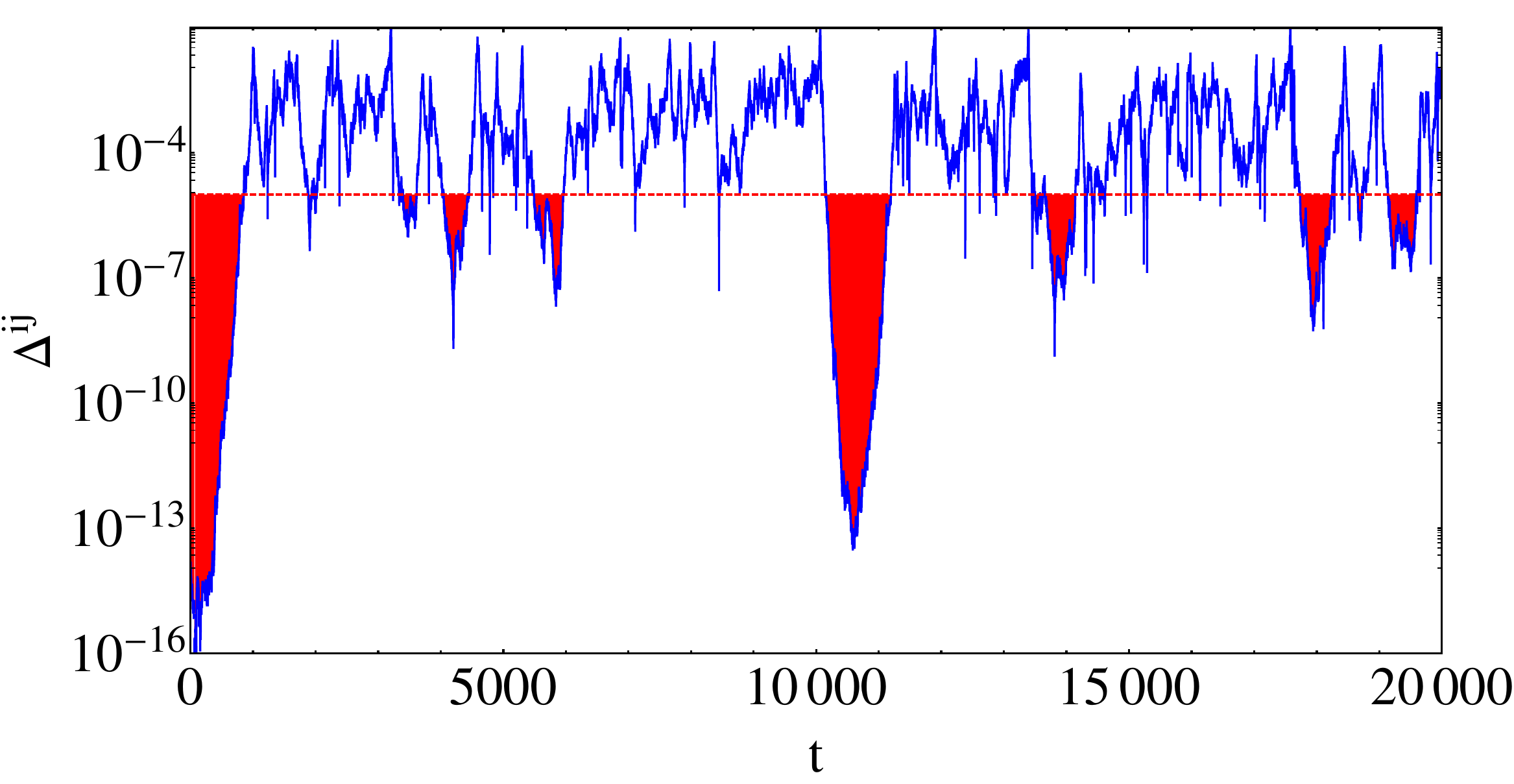}
	\caption{(Color online) Time evolution of the desynchronization $\Delta_{t}^{ij} = \left|x_t^i - x_t^j\right|$ between two units in a completely connected network consisting of $100$ nodes, whose dynamics is determined by \eqref{eq:dynamics} and \eqref{eq:adaptation} with $\tau = - \frac{1}{\text{ln}\left(0.9999\right)} \approx 10^4$ and $\sigma = 2.0$. For the synchronization threshold set to $\Delta_c = 10^{-5}$, the regions with the shaded area under the curve indicate the local synchronization pulses between the two units.}
	\label{fig:desy_twoUnits}
\end{figure}

\begin{figure*}
	\centering
		\includegraphics[width=0.8\textwidth]{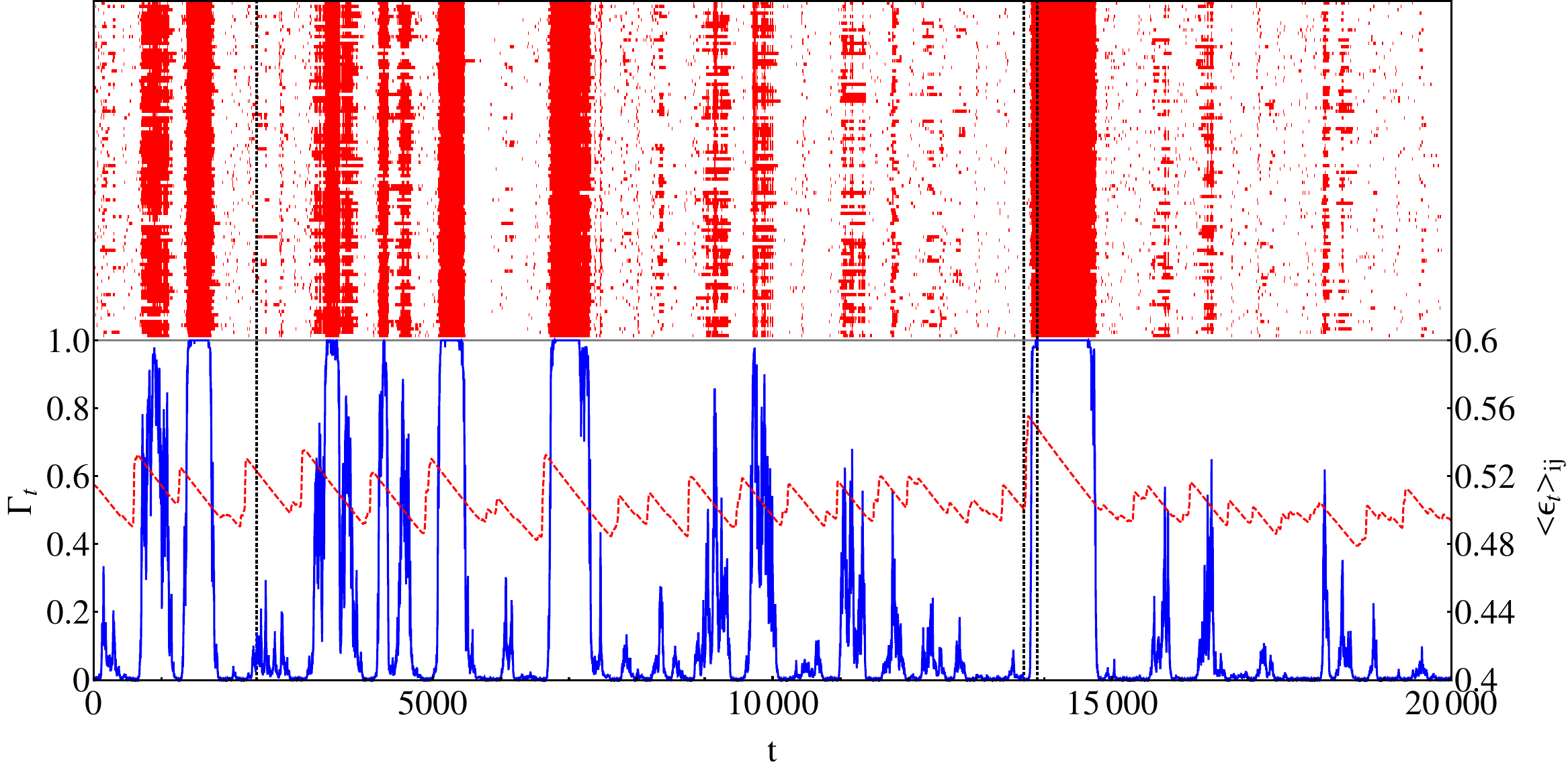}
		\caption{(Color online) Simulations for a completely connected network of $N=100$ nodes. The evolution is determined by the dynamics~\eqref{eq:dynamics} and \eqref{eq:adaptation} with parameters $\tau = - \frac{1}{\text{ln}0.9999} \approx 10^4$ and $\sigma = 2.0$. For an arbitrarily chosen unit $i$, the red lines in the upper part of the plot show the pulses of synchronization with its environment. Every line corresponds to one of the neighbors $j$ of the chosen unit. If the units are synchronized, i.e. $\Delta_t^{ij} < \Delta_c = 10^{-5}$, this is indicated with red in line $j$ at time $t$. In the lower part, the blue curve displays the fraction $\Gamma_t$ of all pairs in the network, which are synchronized at time $t$ (left axis). The red dashed curve shows the coupling strength, averaged over all pairs in the network at each time step (right axis). For the time steps indictated by the vertical dashed lines, the distributions of the coupling strengths are displayed in Fig. \ref{fig:eps}.}
	\label{fig:synch_evolution}
\end{figure*}


\section{Results}

\subsection{Fully connected topology}

The magnitude of the de-synchronization, $\Delta^{ij}_t$, is shown in Fig. \ref{fig:desy_twoUnits} for a single pair $i,j$ of units. The relaxation time is very long, $\tau=10^4$. The units are always highly correlated with each other. But occasionally the coupling strength increases above its static critical value and pulses of synchronization appear.

Below, we are interested in the fraction, $\Gamma$, of bonds which belong to synchronized pairs, in the structure of synchronized clusters, in the strength and duration of pulses, and in the distribution of couplings. All these quantities strongly fluctuate with time; hence, we will numerically calculate their distributions. Throughout this paper, we will present simulations for systems of size $N=100$. However, investigations of systems with up to 800 units did not show any deviating behavior. This is in accordance with the observation that the synchronization border, $\epsilon_c$, for $N=100$ is already close to the thermodynamic limit (Eq. \eqref{eq:border}).

For the simulations, the states of the units are initialized with random values drawn from the interval $\left[0,1\right)$. All couplings are initially set to $\epsilon_{ij}=\epsilon_{ji} = 0.35$ and $\epsilon_{ii}=0$ (no self feedback), i.e. we are below the synchronization threshold for that topology (see Eq. \eqref{eq:border}). Afterwards, the systems are iterated for $10^5$ time steps to make sure that there is no more transient behavior.

\begin{figure}
		\includegraphics[width=0.49\textwidth]{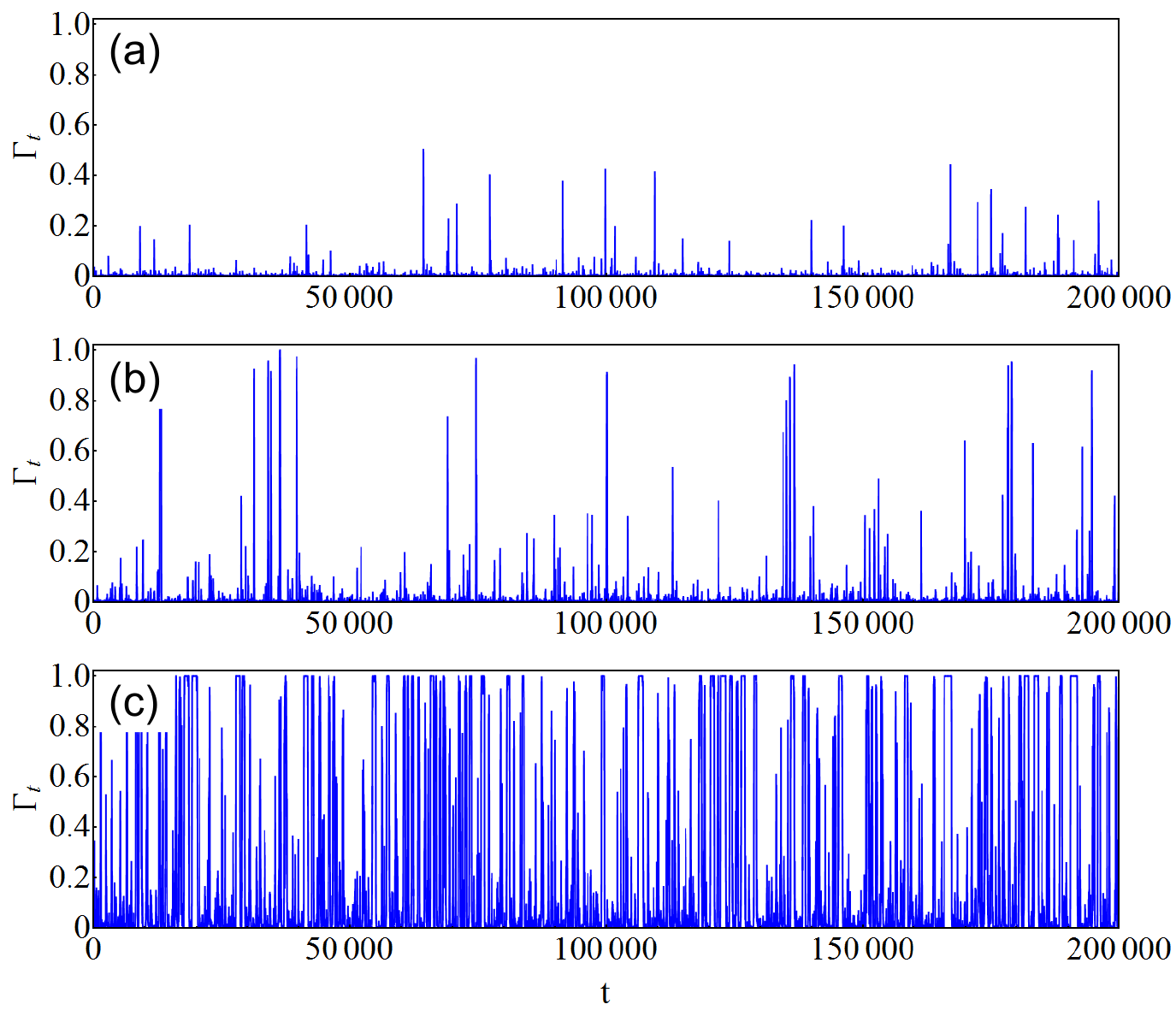}
		\caption{(Color online) Fraction $\Gamma_t$ of all pairs in the network, which are synchronized at time $t$. Simulations for a completely connected network of $N=100$ nodes. The evolution is determined by the dynamics~\eqref{eq:dynamics} and \eqref{eq:adaptation} with parameters $\sigma = 2.0$ and (a) $\tau = - \frac{1}{\text{ln}0.999} \approx 10^3$, (b) $\tau = - \frac{1}{\text{ln}0.9995} \approx 2 \times 10^3$, and (c) $\tau = - \frac{1}{\text{ln}0.9999} \approx 10^4$.}
	\label{fig:synch_evolution_mult}
\end{figure}

Let us define synchronization between two units $i$ and $j$ with a threshold value $\Delta^{ij} < \Delta_c$. Of course, due to the finite range of the state variables, a choice of a too large $\Delta_c$ will cause pairs of units to be mistakenly classified as synchronized. Naturally, in the limit $\Delta_c \to 1$ this is the case for all pairs, even for purely randomly picked state variables. In this work, we set $\Delta_c \equiv 10^{-5}$. For this choice of the threshold, $\Delta_c$, two uniformly distributed randomly chosen values of the variables $x^i$ will lead to false synchronization classification with probability 0.002\%, only. However, the observed behavior is robust under variation of this threshold. According results were found for $\Delta_c = 10^{-4}$.

The upper part of Fig. \ref{fig:synch_evolution} shows the units which are synchronized with unit $x^1$. One sees that synchronization is a cooperative effect; many units contribute to the synchronization pulses. All of the 99 units $x^j$ which are synchronized to unit $x^1$ are shown as dots. From time to time many units are synchronized to the first one, and occasionally even the entire system is synchronized. This is also shown in the lower part of Fig. \ref{fig:synch_evolution}. The fraction, $\Gamma$, of synchronized couplings shows strong irregular peaks. When the complete system is synchronized ($\Gamma=1$), the pulse extends over a large time interval. Then, the average value of the coupling strengths, also shown in Fig. \ref{fig:synch_evolution}, decreases until it falls below the critical coupling, $\epsilon_c$, and the system de-synchronizes. The couplings increase fast and decay slowly in contrast to the fast rise and decay of chaos synchronization. In Fig. \ref{fig:synch_evolution_mult} the fraction $\Gamma(t)$ of pairs which are synchronized is shown as a function of time for different relaxation times $\tau$. With increasing $\tau$ the fraction of synchronized pairs increases and pulses of synchronization become more frequent. Moreover, the duration of the pulses increases.

\begin{figure}
	\centering
		\includegraphics[width=0.49\textwidth]{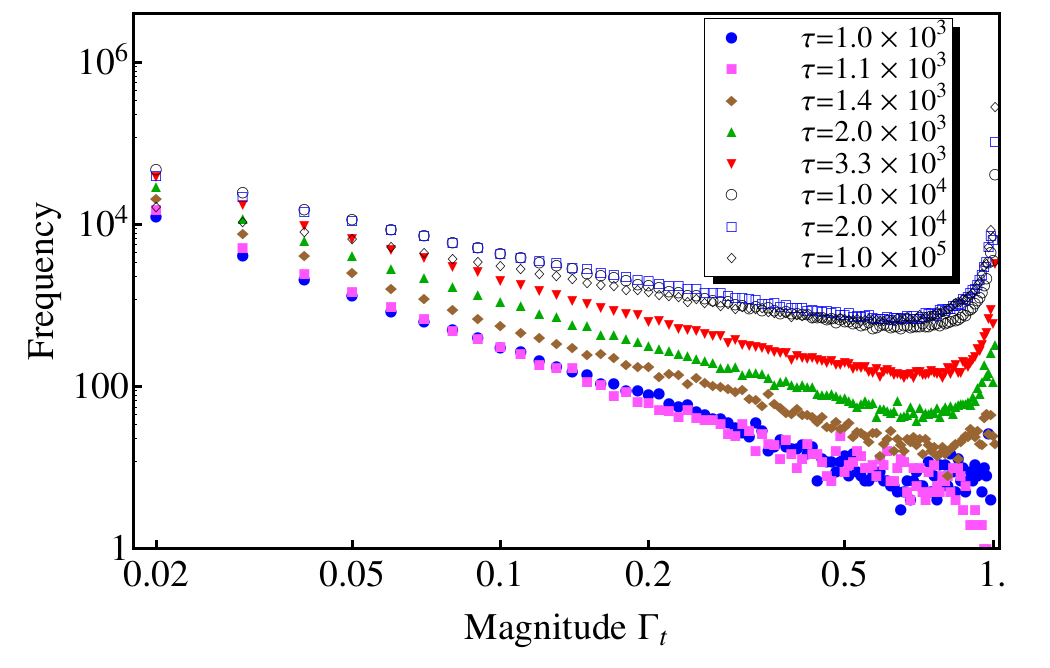}
	\caption{(Color online) Histograms for the occurrence of the heights of the global synchronization $\Gamma$ for a recording time of $5 \times 10^5$ time steps. Each \textit{time step} $t$ contributes with its $\Gamma_t$ value to the frequency i.e. each histogram has $5 \times 10^5$ entries. The bin width is 1\%. Notice the double-logarithmic scale.}
	\label{fig:puls_height}
\end{figure}

Pulses of chaos synchronization occur on all scales. The fraction of synchronized pairs ranges from a small number to complete synchronization of all nodes in the whole system. This is shown in Fig. \ref{fig:puls_height} where the frequency of occurrence of the $\Gamma$-values is plotted as a function of their magnitudes. Each \textit{time step} $t$ contributes with its $\Gamma_t$ value to the frequency.

\begin{figure}
	\includegraphics[width=0.49\textwidth]{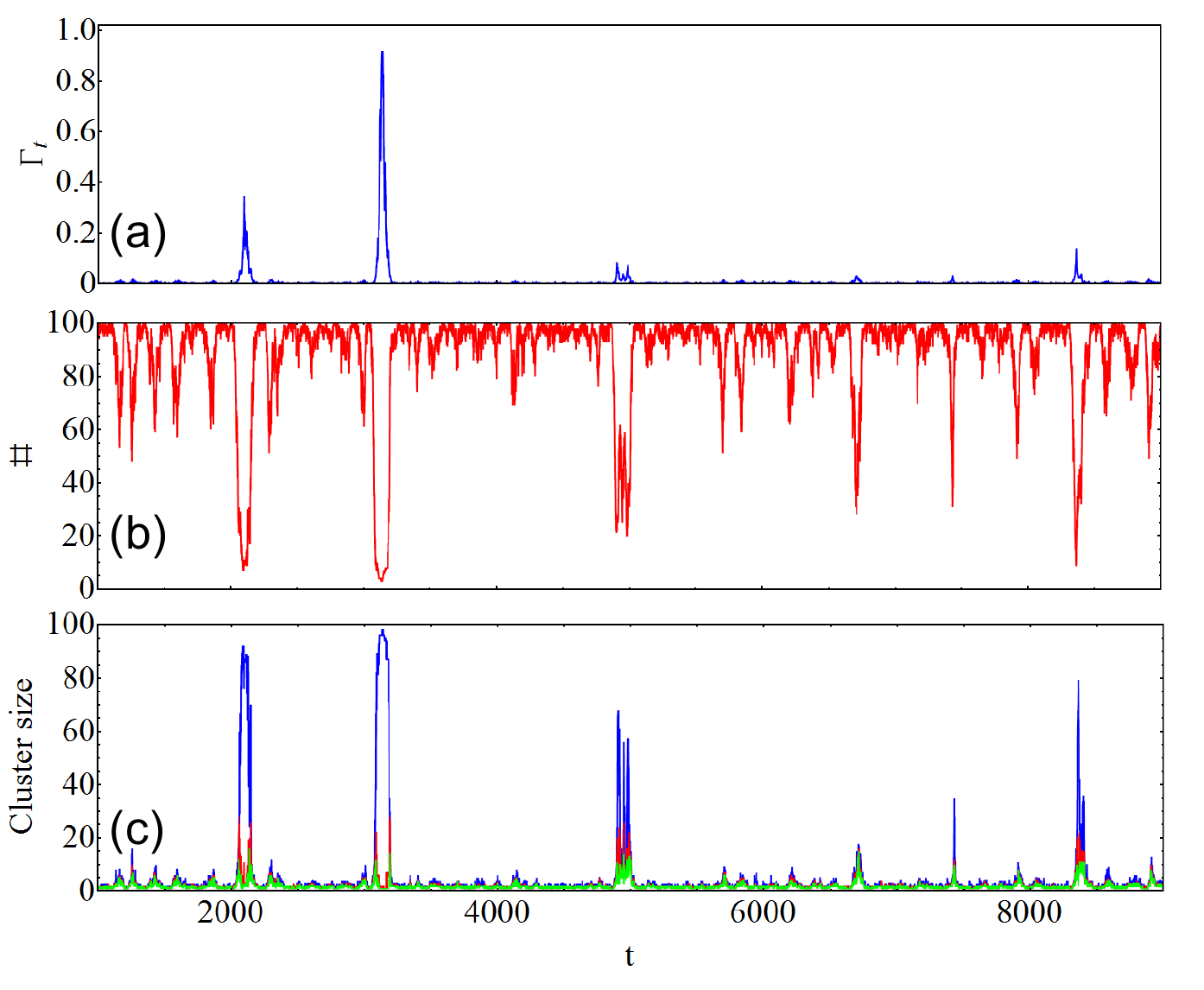}
	\caption{(Color online) Simulations for a completely connected network of $N=100$ nodes. The evolution is determined by the dynamics~\eqref{eq:dynamics} and \eqref{eq:adaptation} with parameters $\sigma = 2.0$ and $\tau = - \frac{1}{\text{ln}0.999} \approx 10^3$. (a): Fraction $\Gamma$ of synchronized pairs in the system. (b): Number of synchronization clusters. (c): Size of the largest three clusters.}
	\label{fig:cluster}
\end{figure}

How is the fraction, $\Gamma$, of synchronized pairs in the network related to the synchronization structure in terms of cluster formation? 
We define a cluster as a set of units which are connected by synchronized couplings (single-link clustering). Fig. \ref{fig:cluster} (b) shows the number of clusters as a function of time. It corresponds to the time evolution of $\Gamma$ displayed in Fig. \ref{fig:cluster} (a). Most of the time the system is not synchronized and the number of clusters is of order of the size $N$ of the network. But occasionally the units cooperate and generate a few clusters. This is also shown in Fig. \ref{fig:cluster} (c) where the size of the three largest clusters is plotted versus time. Sometimes, even the whole network becomes a single cluster with a common chaotic trajectory over several hundreds of time steps. In order to investigate the distributions of the cluster sizes, extensive simulations on larger systems need to be conducted.

\begin{figure}
	\centering
		\includegraphics[width=0.46\textwidth]{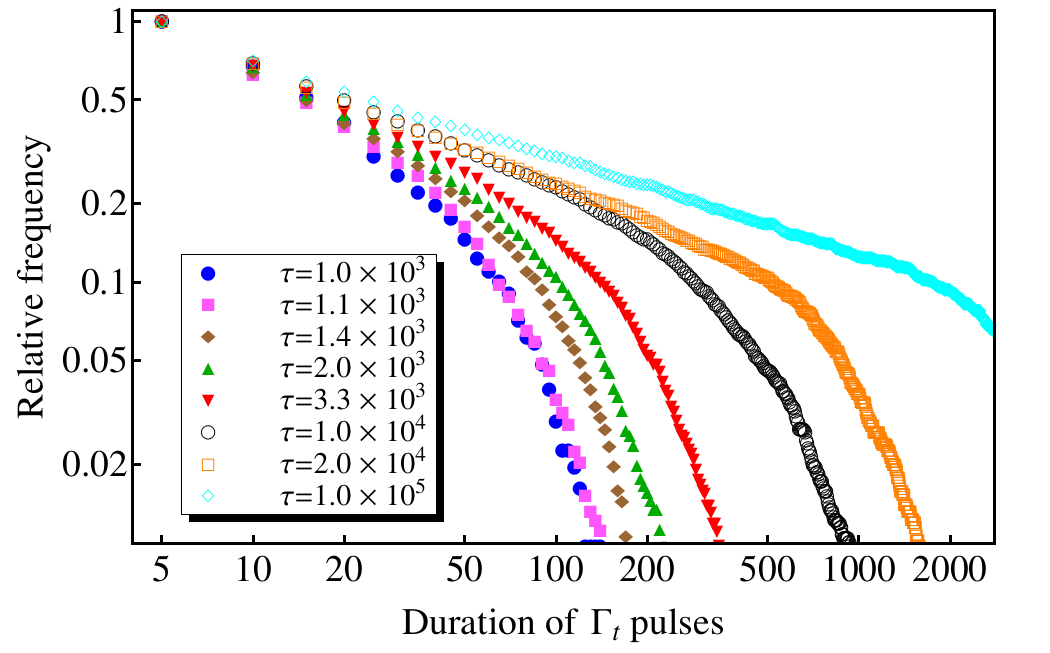}
	\caption{(Color online) Cumulative histograms of the durations of global synchronization pulses for varying $\tau$ and fixed $\sigma = 2.0$. Every entry indicates the relative frequency of pulses with duration equal to or larger than the number of time steps shown on the horizontal axis. The bin width is five time steps. By definition, a global pulse starts, when the global synchronization exceeds the threshold $\Gamma_c = 1\%$, and it ends after falling below the threshold. Notice the double logarithmic scale.}
	\label{fig:duration}
\end{figure}

The duration of the pulses is distributed over a broad range, as well. To supress the influence of the choice of the bin width, Fig. \ref{fig:duration} shows its cumulative histograms. It can be inferred that longer $\Gamma_t$ pulses appear less frequently than shorter ones. Of course, due to the finite recording time of $5 \times 10^5$ time steps, the lower probability for events in the tails of the distributions leads to cutoffs in the latter. This cutoff shifts to larger durations for larger $\tau$. Furthermore, revisiting Fig. \ref{fig:puls_height}, there is a tendency towards more pulses for larger decay times $\tau$; curves corresponding to larger $\tau$ lie above those corresponding to smaller $\tau$. However, this does no longer hold for $\tau > 10^4$. This is because for $\tau > 10^4$, pulses with $\Gamma = 1$ and a length of more than 1000 time steps are observed. Since all distributions are recorded over time frames of equal size ($5 \times 10^5$ time steps), the appearance of very long pulses with $\Gamma = 1$ neccessarily results in a decrease in the number of time steps with $\Gamma < 1$ within the time frame.

\begin{figure}
	\includegraphics[width=0.35\textwidth]{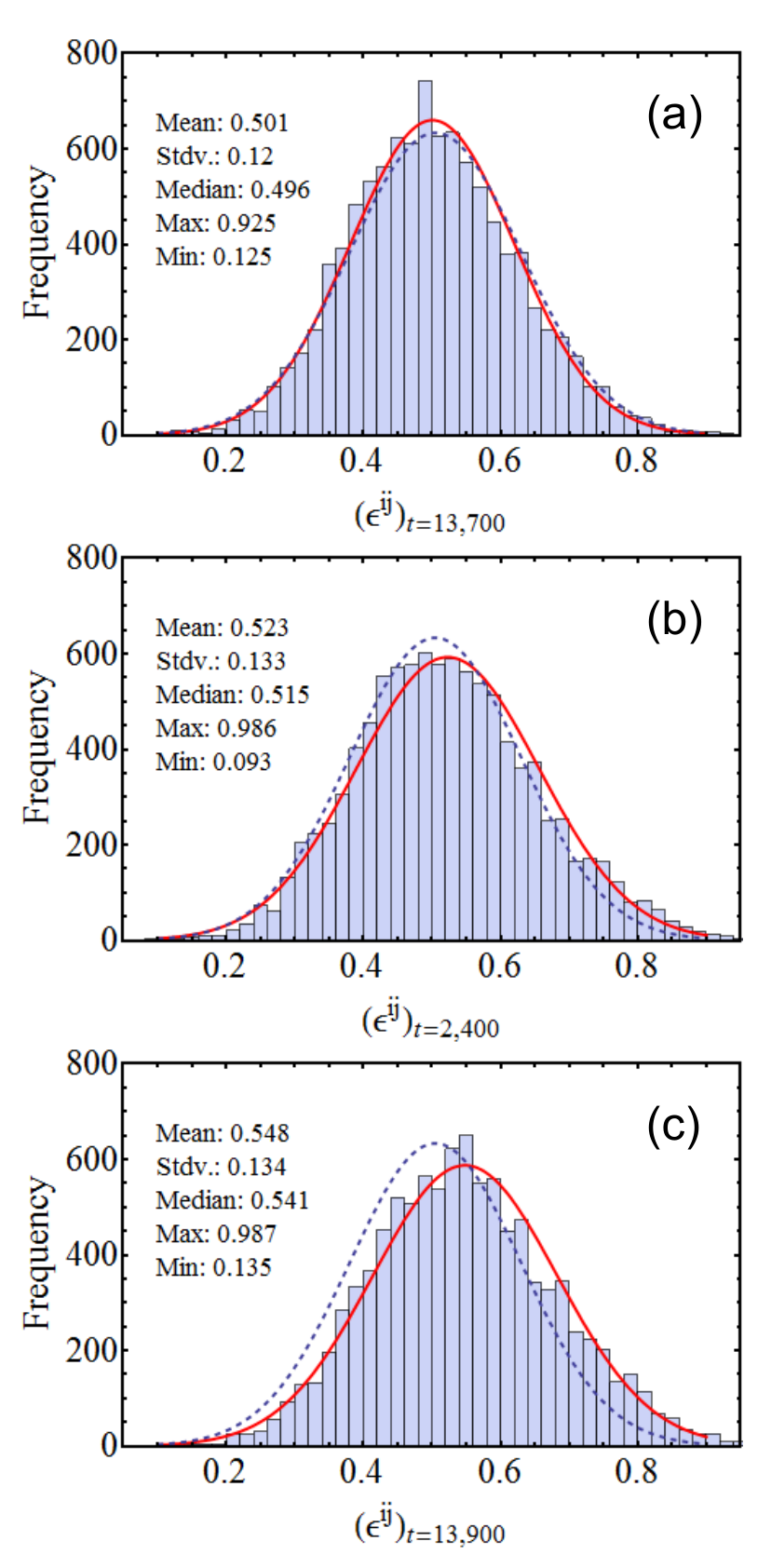}
	\caption{(Color online) Distribution of coupling strengths $\epsilon^{ij}$ for different time steps in Fig. \ref{fig:synch_evolution}. The fits indicated by the red solid lines are Gaussian curves with according mean and standard deviation. The blue dashed lines are Gaussian curves, where the mean and standard deviation are obtained from averages over all couplings for $2 \times 10^5$ time steps.}
	\label{fig:eps}
\end{figure}

The distribution of the couplings is not critical. Fig. \ref{fig:eps} shows a Gaussian-like distribution with a time dependent mean value which fluctuates around the critical value of the static homogeneous network.

\begin{figure}
	\includegraphics[width=0.4\textwidth]{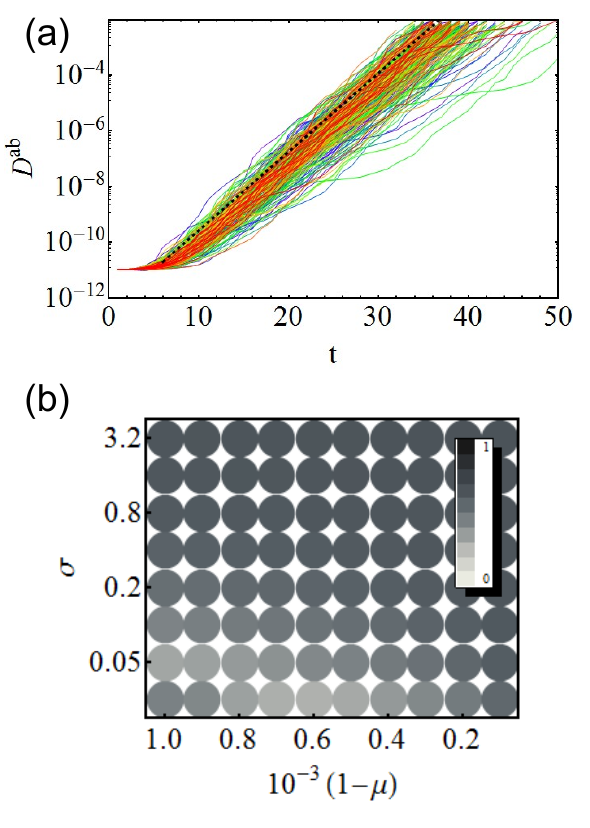}
	\caption{(Color online) Evaluation of the largest Lyapunov exponent for complex systems with multiple variables. The trajectories in (a) show the divergence $\mathcal{D}^{ab}_t$ of a system $a$, with $\tau = 10^3$ and $\sigma = 2.0$, and its cloned version $b$ in which every variable is perturbed by a value of $10^{-15}$. Once the distance exceeds a value of $10^{-2}$, the copy is reinitialized and $t$ is reset to zero. Logarithmic fits of the data yield a distribution of Lyapunov exponents. The slope of the dashed line in (a) corresponds to the averaged Lyapunov exponent over all fits. It is the largest Lyapunov exponent (LLE) of the system. Fig. (b) shows the LLE for all investigated parameter sets $\left(\mu,\sigma\right)$.}
	\label{fig:LLE}
\end{figure}

The activity of the network is rather irregular, but is it still chaotic? We calculated the difference between two nearby trajectories $a$ and $b$ in the space of the dynamic variables and coupling strengths, defined by 
\begin{equation} \label{eq:LLE}
	\begin{split}
		\mathcal{D}^{ab}_t &= \sum_{i=1}^N \left[\left|x_t^{i \, \left(a\right)} - x_t^{i \,\left(b\right)}\right|  \, + \,  \sum_{j=1,j\neq i}^N \left|\epsilon_t^{ij \, \left(a\right)} - \epsilon_t^{ij \,\left(b\right)}\right|  \right].
	\end{split}
\end{equation}

A log-linear plot of $D^{ab}_t$ clearly shows an exponential divergence (see Fig. \ref{fig:LLE} (a)) with a positive, but small  Lyapunov exponent for all parameters $\tau$ and $\sigma$ which we investigated (see Fig. \ref{fig:LLE} (b)). Thus, the dynamics of the systems are still chaotic.

\begin{figure*}
	\centering
		\includegraphics[width=0.9\textwidth]{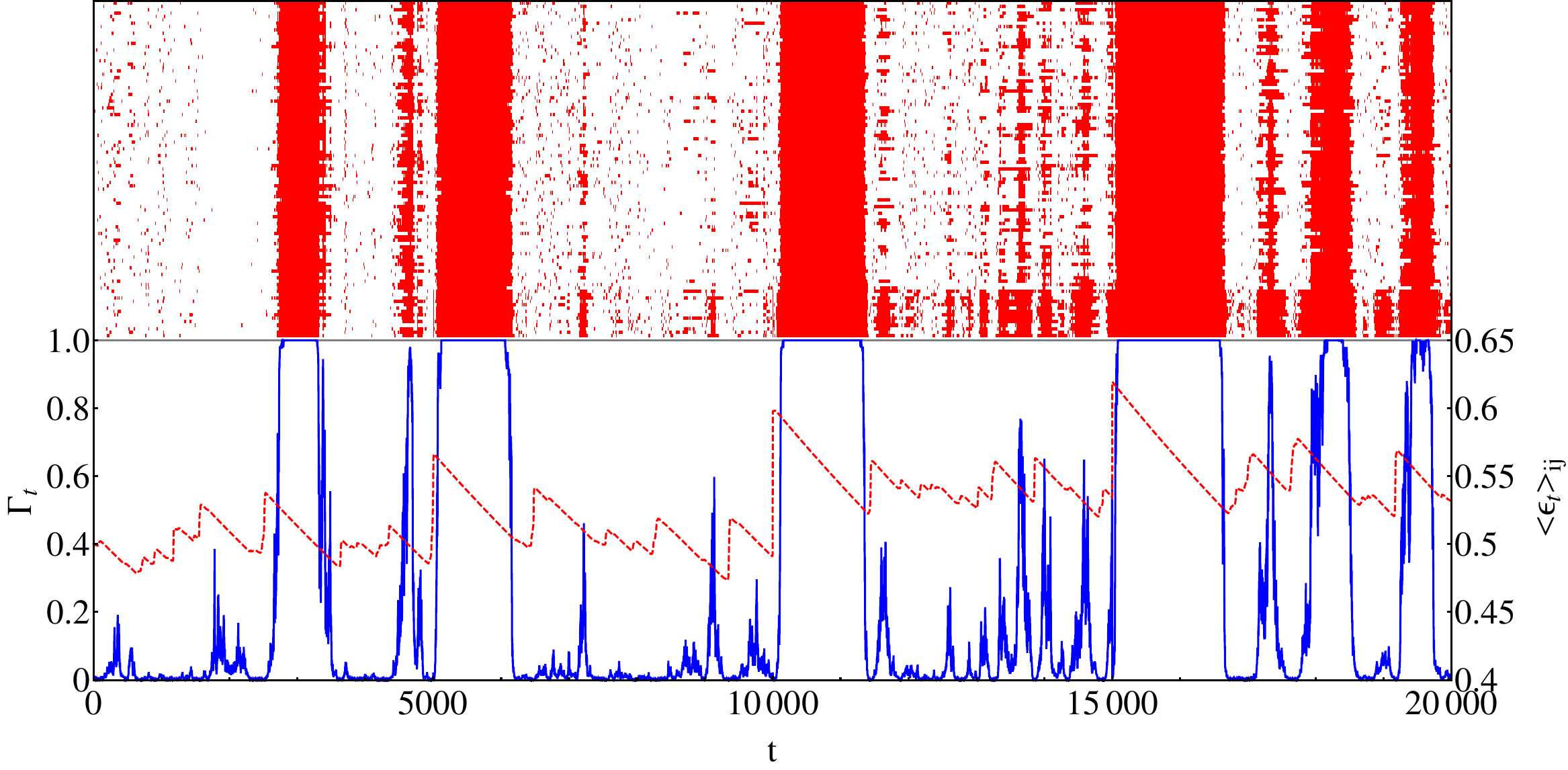}
		\caption{(Color online) Simulations for a completely connected network of N=100 nodes. The evolution is determined by the dynamics~\eqref{eq:dynamics} and \eqref{eq:adaptation} with parameters $\tau = - \frac{1}{\text{ln}0.9999} \approx 10^4$ and $\sigma = 2.0$. At time steps 5,000, 10,000, and 15,000, clusters of $N_{\text{stim}} = 15$ units, and thus $\Gamma_{\text{stim}} = \frac{N_{\text{stim}} \left(N_{\text{stim}}-1\right)}{N \left(N-1\right)} \times 100\% \approx 2.12\%$ of the pairs, are stimulated in terms of both their states and their mutual couplings. The red pulses in the upper part indicate the local synchronization of one unit of the stimulated cluster with the remaining 99 units. Among them, the lower $\left(N_{\text{stim}} - 1\right)$ lines indicate the synchronization with the other stimulated nodes. In the lower part, the blue lines display the trajectories of $\Gamma_t$, and the red dashed curves indicate the mean coupling strengths at time $t$, averaged over all pairs of units in the systems.}
	\label{fig:synch_evolution_stim}
\end{figure*}

For biological applications, the sensitivity of the networks to external stimuli is of particular interest. We investigate the systems' responses to excitations of the following form. A stimulus is realized by, firstly, imposing the state
\begin{equation} \label{eq:stimNodes}
	x^{i}_{\text{stim}} = X_{\text{stim}} + \delta X^{i}_{\text{stim}}
\end{equation}
to the first $N_{\text{stim}}$ nodes $i \in \left\{1, \ldots , N_{\text{stim}}\right\}$, where $X_{\text{stim}}$ and $\delta X^{i}_{\text{stim}}$ are uniformly distributed random numbers with $X_{\text{stim}} \in \left[0\,,\,0.99\right)$ and $\delta X^{i}_{\text{stim}} \in \left[0\,,\,10^{-10}\right)$. The mutual synchronization of the $N_{\text{stim}}$ nodes contributes to $\Gamma$ with
\begin{equation} 
	\Gamma_{\text{stim}} = \frac{\binom{N_{\text{stim}}}{2}}{\binom{N}{2}} = \frac{N_{\text{stim}} \left(N_{\text{stim}}-1\right)}{N \left(N-1\right)}.
\end{equation}
In addition, their mutual couplings are set to the maximum magnitude of one,
\begin{equation} \label{eq:stimCouplings}
	\epsilon^{ij}_{\text{stim}} = 1\,\, \, \, \forall \, \, \, i,j \neq i \in \left\{1, \ldots , N_{\text{stim}}\right\}.
\end{equation}
Thus, here a stimulus means that an almost completely synchronized state is imposed on the cluster of $N_{\text{stim}}$ excited nodes and the corresponding nodes are maximally coupled. Can a small fraction of synchronized units drive the whole network to complete synchronization before their mutual couplings have decayed below the synchronization threshold? Fig. \ref{fig:synch_evolution_stim} shows simulation results of a system with $\tau = 10^4$ in which $N_{\text{stim}} = 15$ units and $\Gamma_{\text{stim}} \approx 2.12\%$ of the couplings are stimulated at time steps 5,000, 10,000, and 15,000. Beside the naturally occurring synchronization pulses, every applied stimulus leads to a pulse in which the entire network becomes synchronized. If the number of stimulated units is smaller than a critical threshold, the network will not respond to the stimulus. The fraction of the network that needs to be stimulated to trigger pulses which extend to the whole system depends on the decay time, $\tau$ of the couplings. The upper part of Fig. \ref{fig:synch_evolution_stim} shows the synchronization of one stimulated unit with its environment, while the lower 14 lines correspond to the other stimulated units. We infer that even several thousand time steps after the excitations, there is an increased probability for the stimulated cluster to synchronize, i.e. the system memorizes the stimuli over the time scale $\tau$ of the couplings.


\subsection{2-d lattice with long-range interactions}

\begin{figure}
	\centering
		\includegraphics[width=0.4\textwidth]{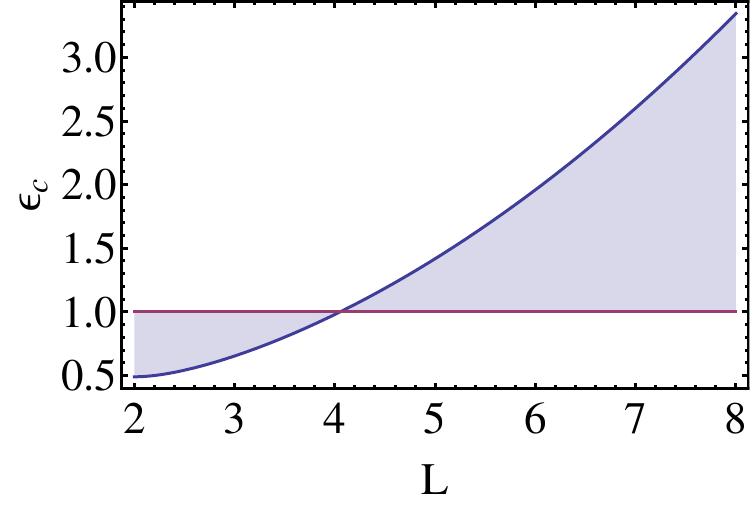}
		\caption{(Color online) Synchronization threshold $\epsilon_c$ for a two-dimensional square lattice (p=0) with uniform coupling strengths, depending on the lateral system size $L = \sqrt{N}$.}
	\label{fig:synch_threshold_p0}
\end{figure}

\begin{figure}
	\includegraphics[width=0.5\textwidth]{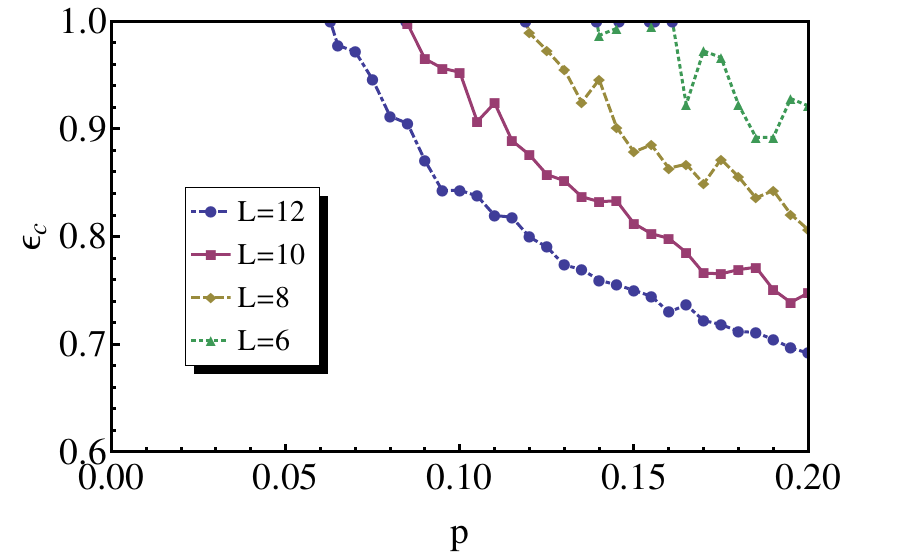}
	\caption{(Color online) Synchronization threshold $\epsilon_c$ for a network generated from a two-dimensional grid, depending on the probability $p$ of establishing additional connections for several system sizes $N=L^2$. Every point is an average over five topologies with the same value of $p$. $\epsilon_c$ can be calculated from the spectral properties of the coupling matrix 
\eqref{eq:couplingThresholdGeneral}.}
	\label{fig:thresh}
\end{figure}

In addition to fully connected systems we also considered two-dimensional square lattices with nearest neighbor couplings. To suppress finite-size effects the boundaries are continued periodically. Square lattices with lateral size $L$ and uniform coupling strengths are not able to entirely synchronize for $L > 4$ because their hypothetical \footnote{Since $\epsilon \in \left[0,1\right]$, synchronization thresholds $\epsilon_c > 1$ are purely hypothetical, although they can be calculated from the eigenvalues of the coupling matrix. They indicate that the correpsonding systems may not entirely synchronize.} synchronization threshold, $\epsilon_c$, is larger than one, as will be shown below. Numerical investigations indicate that this is neither possible for an adaptation of the couplings according to \eqref{eq:adaptation}. Therefore, we add long-range couplings to the short-range interactions of the square lattice. By randomly introducing bidirectional links to the system with probability $p$, the hypothetical synchronization threshold, determined by the graph spectrum of the static network, decreases. Thus, for sufficiently large $p$, $\epsilon_c$ falls below one and the networks with static uniformly weighted couplings can be synchronized (Fig. \ref{fig:thresh}).

We will investigate two-dimensional grids with additional long-range links and adaptive coupling weights. Since the node degree $k^i$ is no longer $N-1$ for all vertices, dynamics \eqref{eq:dynamics} will be modified to
\begin{equation} \label{eq:dynamics_grid}
  x_{t}^{i} = \left(1-\sum_{j\neq i}\frac{\epsilon_{t-1}^{ij} \, A^{ij}}{k^i}\right)\,f(x_{t-1}^{i})+\sum_{j\neq i}\frac{\epsilon_{t-1}^{ij} \, A^{ij}}{k^i}\,f(x_{t-1}^{j})
\end{equation} 
with the adjacency matrix \textbf{A} \footnote{The adjacency matrix \textbf{A} has entries $A^{ij} = 1$ if there is a link from unit $j$ to unit $i$. All other entries are zero.}. In analogy to \eqref{eq:couplingThreshold}, the criterion for the stability of a system with static uniform couplings, $\epsilon_t^{ij} = \epsilon$, is given by
\begin{equation} \label{eq:couplingThresholdGeneral}
	\frac{1-e^{-\lambda}}{1-\gamma_1} \equiv \epsilon_c < \epsilon < \frac{1+e^{-\lambda}}{1-\gamma_1},
\end{equation}
with $\gamma_1$ being the second largest eigenvalue of the coupling matrix $\boldsymbol{G}$, which equals the adjacency matrix with the row sums normalized to unity: $G^{ij} \equiv \frac{1}{k^{i}} A^{ij}$.

\begin{figure}
	\centering
		\includegraphics[width=0.50\textwidth]{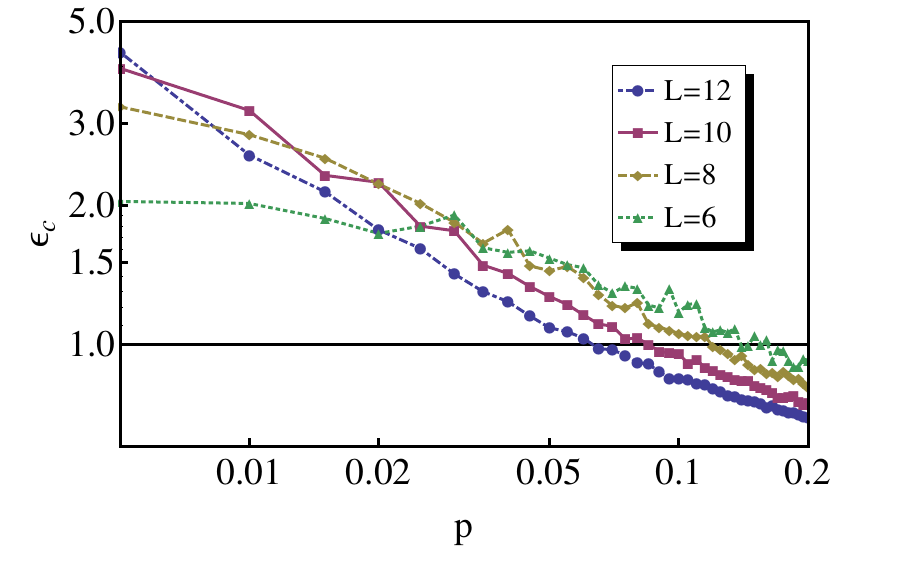}
	\caption{(Color online) Double-logarithmic display of the critical coupling strengths of Fig. \ref{fig:thresh}. The illustration is extended to the theoretically calculated values with magnitude larger than one.}
	\label{fig:synch_threshold_loglog}
\end{figure}

\begin{figure}
		\includegraphics[width=0.49\textwidth]{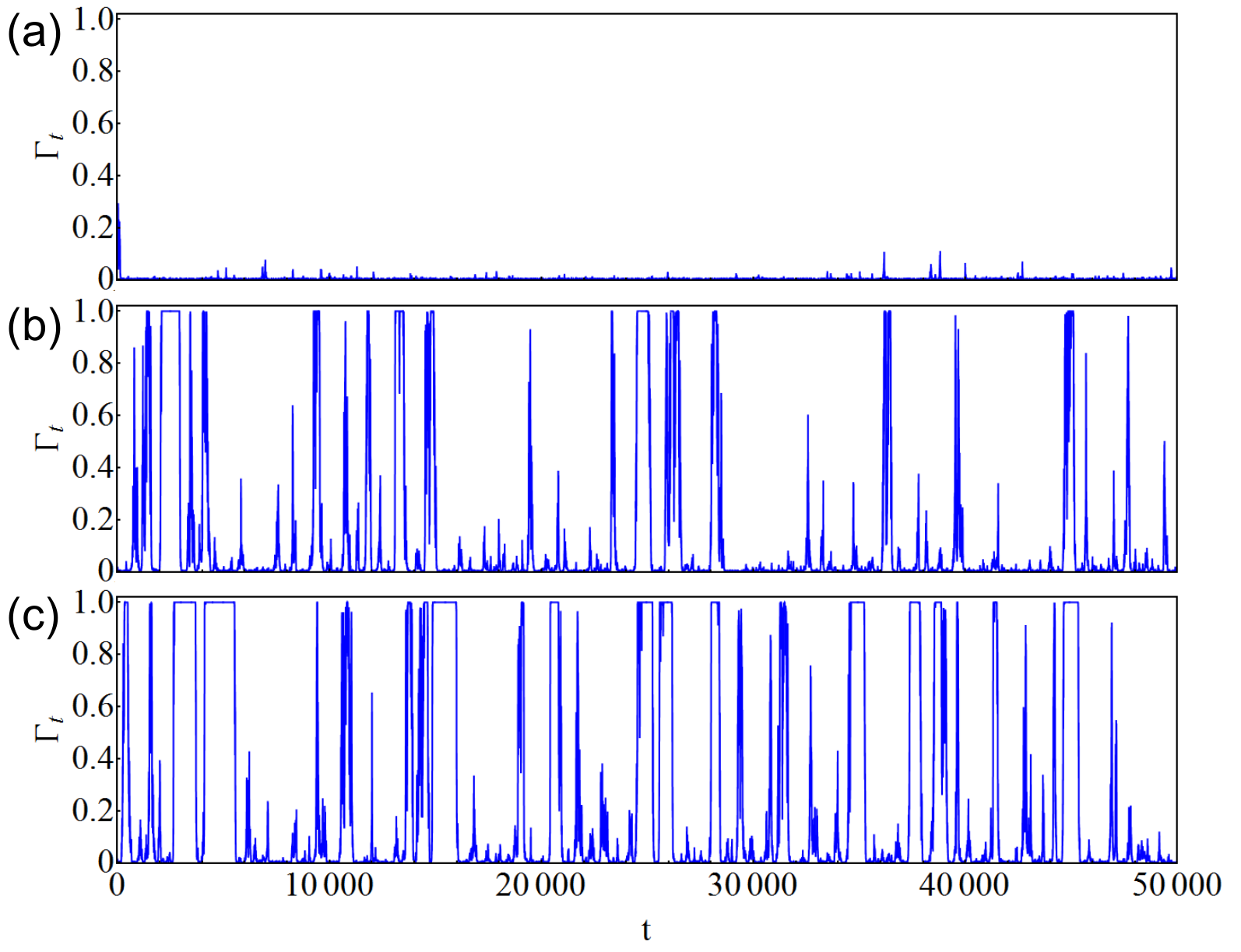}
		\caption{(Color online) Fraction $\Gamma_t$ of all pairs in the network, which are synchronized at time $t$. Simulations for a two-dimensional square lattice with additional long-range interactions with $N=L^2=100$ nodes. The evolution is determined by the dynamics~\eqref{eq:dynamics_grid} and \eqref{eq:adaptation} with parameters $\tau = - \frac{1}{\text{ln}0.9999} \approx 10^4$ and $\sigma = 2.0$. Long-range couplings are added with probability (a) $p=0.08$, (b) $p=0.15$, and (c) $p=0.4$.}
	\label{fig:synch_evolution_grid}
\end{figure}

For two-dimensional hypercubic grids of linear size $L$ with uniform coupling strenghts (i.e. the coupling strength between all pairs of neighboring units $i$ and $j$ is $G^{ij} = \frac{1}{4}$) and periodic boundary conditions, it is known that the eigenvalues $\lambda_k$ of the Laplacian matrix $\boldsymbol{L} = 4 \, \left(\mathds{1} - \boldsymbol{G}\right)$ are given by
\begin{equation}
	\begin{split}
	\lambda_{\left\{m_{\alpha}\right\}} = 4 \, \left( \text{sin}^2 \frac{\pi m_{1}}{L} + \text{sin}^2 \frac{\pi m_{2}}{L} \right)
	\end{split}
\end{equation}
where $\left\{m_\alpha\right\}$ is a $2$-tuple with $m_\alpha \in \left\{0,1, \dots , L-1\right\}$ and $L^2 = N$ (see \cite{hongKimChoiPark}). 
Since \textbf{L} and \textbf{G} have the same eigenvectors, the relation between $\gamma_k$ and $\lambda_k$ is given by
\begin{equation}
	\gamma_k = 1 - \frac{\lambda_k}{4}.
\end{equation}
Consequently, the second largest $\gamma_k$ corresponds to the second smallest $\lambda_k$. The latter is given by
\begin{equation}
	\lambda_1 = \lambda_{01} = \lambda_{10} = \lambda_{0(L-1)} = \lambda_{(L-1)0} = 4 \, \text{sin}^2 \frac{\pi}{L}.
\end{equation}
Hence, for a 2-dimensional grid equation \eqref{eq:couplingThresholdGeneral} yields
\begin{equation}
	\frac{1-e^{-\lambda}}{\text{sin}^2 \frac{\pi}{L}} \equiv \epsilon_c < \epsilon < \frac{1+e^{-\lambda}}{\text{sin}^2 \frac{\pi}{L}}.
\end{equation}

Fig. \ref{fig:synch_threshold_p0} shows the critical coupling strength, $\epsilon_c$, versus the lateral system size $L$. For $L > 4$ it exceeds one and thus no complete synchronization is possible. While this result is derived from grids with uniform coupling strenghts, numerical simulations reveal that even for an adaptive variation of the couplings no complete synchronization can be achieved.

The fact that complete synchronization is not possible for $L > 4$ seems to contradict the observation that larger systems with a certain fraction $p$ of long-range interactions posses a tendency towards lower synchronization thresholds $\epsilon_c$ (Fig. \ref{fig:thresh}). However, Fig. \ref{fig:synch_threshold_loglog} shows an extended plot range of the numerically calculated synchronization thresholds, including hypothetical values of $\epsilon_c > 1$. One finds that as $p$ approaches zero, the curves intersect at some point and for $p=0$ those systems with larger lateral size exhibit a larger $\epsilon_c$.

Fig. \ref{fig:synch_evolution_grid} shows simulation results of the dynamics \eqref{eq:dynamics_grid} for networks with $\tau = 10^4$, $\sigma = 2.0$, and $N = L^2 = 100$ for various $p$. Up to $p \lesssim 0.07$ we do not observe synchronization between the units. With $p \gtrsim 0.08$ ($L=10$) the synchronization threshold of a static system, in Fig. \ref{fig:thresh} drops below one which makes synchronization possible. Above $p \geq 0.08$ we also find pulses of synchronization in the network with adaptive couplings. Both the duration and intensity of the $\Gamma$-spikes increase with $p$ as the synchronization threshold of the corresponding static network decreases (Fig. \ref{fig:synch_evolution_grid}). For $p = 0.4$ the results in figure \ref{fig:synch_evolution_grid} (c) resemble those for a completely connected topology. According results were found for $\tau = 10^3$.

As for the completely connected systems, both the magnitude of $\Gamma$ and the duration of the $\Gamma$ pulses follow heavy-tailed distributions.


\section{Discussion}

We have investigated a coupled map lattice which shows chaos, synchronization, and criticality. These macroscopic properties are generated by dynamic couplings with competing adaptation rules. The network relaxes to a chaotic state which exhibits pulses of chaos synchronization. Synchronized activity has been observed for all sizes as well as for all time scales with corresponding heavy-tailed distributions. An all-to-all coupled network possesses similar properties as a square lattice with additional long-range couplings. The network memorizes a stimulus for a long period if the size of stimulated units is large enough.

Pulsed chaos synchronization has been shown by numerical simulations of coupled map lattices. However, we expect that synchronized pulses will be found in networks of chaotic differential equations, as well; but this has to be shown.  

An interesting question is whether pulsed synchronization of an irregular dynamics is a relevant mechanism of biological neural networks. Synchronization is believed to be involved in the transmission and processing of information in the cortex. Here we consider a homogenous network of nonlinear units with irregular dynamics. In the resting state, i.e. without stimulus and without learning, the units may communicate with each other by spontaneously creating clusters of all sizes. On top of the static topological network a dynamic network of synchronized clusters emerges. These clusters can be triggered by a small group of synchronized units. Hence, for biological applications, one might extend this research in the direction of more realistic models, from integrate-and-fire to Hodgkin-Huxley models, consider different time scales for the synaptic plasticity compared to the neuronal dynamics, and investigate the response to stimuli and the effect of learning in more detail.

\bibliographystyle{apsrev4-1}

\begin{thebibliography}{30}%
\makeatletter
\providecommand \@ifxundefined [1]{%
 \@ifx{#1\undefined}
}%
\providecommand \@ifnum [1]{%
 \ifnum #1\expandafter \@firstoftwo
 \else \expandafter \@secondoftwo
 \fi
}%
\providecommand \@ifx [1]{%
 \ifx #1\expandafter \@firstoftwo
 \else \expandafter \@secondoftwo
 \fi
}%
\providecommand \natexlab [1]{#1}%
\providecommand \enquote  [1]{``#1''}%
\providecommand \bibnamefont  [1]{#1}%
\providecommand \bibfnamefont [1]{#1}%
\providecommand \citenamefont [1]{#1}%
\providecommand \href@noop [0]{\@secondoftwo}%
\providecommand \href [0]{\begingroup \@sanitize@url \@href}%
\providecommand \@href[1]{\@@startlink{#1}\@@href}%
\providecommand \@@href[1]{\endgroup#1\@@endlink}%
\providecommand \@sanitize@url [0]{\catcode `\\12\catcode `\$12\catcode
  `\&12\catcode `\#12\catcode `\^12\catcode `\_12\catcode `\%12\relax}%
\providecommand \@@startlink[1]{}%
\providecommand \@@endlink[0]{}%
\providecommand \url  [0]{\begingroup\@sanitize@url \@url }%
\providecommand \@url [1]{\endgroup\@href {#1}{\urlprefix }}%
\providecommand \urlprefix  [0]{URL }%
\providecommand \Eprint [0]{\href }%
\providecommand \doibase [0]{http://dx.doi.org/}%
\providecommand \selectlanguage [0]{\@gobble}%
\providecommand \bibinfo  [0]{\@secondoftwo}%
\providecommand \bibfield  [0]{\@secondoftwo}%
\providecommand \translation [1]{[#1]}%
\providecommand \BibitemOpen [0]{}%
\providecommand \bibitemStop [0]{}%
\providecommand \bibitemNoStop [0]{.\EOS\space}%
\providecommand \EOS [0]{\spacefactor3000\relax}%
\providecommand \BibitemShut  [1]{\csname bibitem#1\endcsname}%
\let\auto@bib@innerbib\@empty
\bibitem [{\citenamefont {Boccaletti}\ \emph {et~al.}(2006)\citenamefont
  {Boccaletti}, \citenamefont {Latora}, \citenamefont {Moreno}, \citenamefont
  {Chavez},\ and\ \citenamefont {Hwang}}]{Boccaletti2006175}%
  \BibitemOpen
  \bibfield  {author} {\bibinfo {author} {\bibfnamefont {S.}~\bibnamefont
  {Boccaletti}}, \bibinfo {author} {\bibfnamefont {V.}~\bibnamefont {Latora}},
  \bibinfo {author} {\bibfnamefont {Y.}~\bibnamefont {Moreno}}, \bibinfo
  {author} {\bibfnamefont {M.}~\bibnamefont {Chavez}}, \ and\ \bibinfo {author}
  {\bibfnamefont {D.-U.}\ \bibnamefont {Hwang}},\ }\href {\doibase
  10.1016/j.physrep.2005.10.009} {\bibfield  {journal} {\bibinfo  {journal}
  {Physics Reports}\ }\textbf {\bibinfo {volume} {424}},\ \bibinfo {pages} {175
  } (\bibinfo {year} {2006})}\BibitemShut {NoStop}%
\bibitem [{\citenamefont {{Arenas}}\ \emph {et~al.}(2008)\citenamefont
  {{Arenas}}, \citenamefont {{D{\'\i}az-Guilera}}, \citenamefont {{Kurths}},
  \citenamefont {{Moreno}},\ and\ \citenamefont {{Zhou}}}]{ArenasReview}%
  \BibitemOpen
  \bibfield  {author} {\bibinfo {author} {\bibfnamefont {A.}~\bibnamefont
  {{Arenas}}}, \bibinfo {author} {\bibfnamefont {A.}~\bibnamefont
  {{D{\'\i}az-Guilera}}}, \bibinfo {author} {\bibfnamefont {J.}~\bibnamefont
  {{Kurths}}}, \bibinfo {author} {\bibfnamefont {Y.}~\bibnamefont {{Moreno}}},
  \ and\ \bibinfo {author} {\bibfnamefont {C.}~\bibnamefont {{Zhou}}},\ }\href
  {\doibase 10.1016/j.physrep.2008.09.002} {\bibfield  {journal} {\bibinfo
  {journal} {Physical Reports}\ }\textbf {\bibinfo {volume} {469}},\ \bibinfo
  {pages} {93} (\bibinfo {year} {2008})}\BibitemShut {NoStop}%
\bibitem [{\citenamefont {Fries}(2005)}]{Fries2005474}%
  \BibitemOpen
  \bibfield  {author} {\bibinfo {author} {\bibfnamefont {P.}~\bibnamefont
  {Fries}},\ }\href {\doibase DOI: 10.1016/j.tics.2005.08.011} {\bibfield
  {journal} {\bibinfo  {journal} {Trends in Cognitive Sciences}\ }\textbf
  {\bibinfo {volume} {9}},\ \bibinfo {pages} {474 } (\bibinfo {year}
  {2005})}\BibitemShut {NoStop}%
\bibitem [{\citenamefont {Womelsdorf}\ \emph {et~al.}(2007)\citenamefont
  {Womelsdorf}, \citenamefont {Schoffelen}, \citenamefont {Oostenveld},
  \citenamefont {Singer}, \citenamefont {Desimone}, \citenamefont {Engel},\
  and\ \citenamefont
  {Fries}}]{Womelsdorf_Schoffelen_Oostenveld_Singer_Desimone_Engel_Fries_2007}%
  \BibitemOpen
  \bibfield  {author} {\bibinfo {author} {\bibfnamefont {T.}~\bibnamefont
  {Womelsdorf}}, \bibinfo {author} {\bibfnamefont {J.-M.}\ \bibnamefont
  {Schoffelen}}, \bibinfo {author} {\bibfnamefont {R.}~\bibnamefont
  {Oostenveld}}, \bibinfo {author} {\bibfnamefont {W.}~\bibnamefont {Singer}},
  \bibinfo {author} {\bibfnamefont {R.}~\bibnamefont {Desimone}}, \bibinfo
  {author} {\bibfnamefont {A.~K.}\ \bibnamefont {Engel}}, \ and\ \bibinfo
  {author} {\bibfnamefont {P.}~\bibnamefont {Fries}},\ }\href
  {http://www.ncbi.nlm.nih.gov/pubmed/17569862} {\bibfield  {journal} {\bibinfo
   {journal} {Science}\ }\textbf {\bibinfo {volume} {316}},\ \bibinfo {pages}
  {1609} (\bibinfo {year} {2007})}\BibitemShut {NoStop}%
\bibitem [{\citenamefont {Buzs{\'a}ki}(2006)}]{buzsaki2006rhythms}%
  \BibitemOpen
  \bibfield  {author} {\bibinfo {author} {\bibfnamefont {G.}~\bibnamefont
  {Buzs{\'a}ki}},\ }\href {http://books.google.com/books?id=ldz58irprjYC}
  {\emph {\bibinfo {title} {Rhythms of the brain}}}\ (\bibinfo  {publisher}
  {Oxford University Press},\ \bibinfo {year} {2006})\BibitemShut {NoStop}%
\bibitem [{\citenamefont {Dehaene}\ and\ \citenamefont
  {Changeux}(2011)}]{Dehaene2011200}%
  \BibitemOpen
  \bibfield  {author} {\bibinfo {author} {\bibfnamefont {S.}~\bibnamefont
  {Dehaene}}\ and\ \bibinfo {author} {\bibfnamefont {J.-P.}\ \bibnamefont
  {Changeux}},\ }\href {\doibase 10.1016/j.neuron.2011.03.018} {\bibfield
  {journal} {\bibinfo  {journal} {Neuron}\ }\textbf {\bibinfo {volume} {70}},\
  \bibinfo {pages} {200 } (\bibinfo {year} {2011})}\BibitemShut {NoStop}%
\bibitem [{\citenamefont {Beggs}\ and\ \citenamefont
  {Plenz}(2003)}]{Beggs2003}%
  \BibitemOpen
  \bibfield  {author} {\bibinfo {author} {\bibfnamefont {J.~M.}\ \bibnamefont
  {Beggs}}\ and\ \bibinfo {author} {\bibfnamefont {D.}~\bibnamefont {Plenz}},\
  }\href {http://www.jneurosci.org/cgi/content/abstract/23/35/11167} {\bibfield
   {journal} {\bibinfo  {journal} {J. Neurosci.}\ }\textbf {\bibinfo {volume}
  {23}},\ \bibinfo {pages} {11167} (\bibinfo {year} {2003})}\BibitemShut
  {NoStop}%
\bibitem [{\citenamefont {Plenz}(2010)}]{Plenz_2010}%
  \BibitemOpen
  \bibfield  {author} {\bibinfo {author} {\bibfnamefont {D.}~\bibnamefont
  {Plenz}},\ }\href {http://www.nature.com/doifinder/10.1038/nphys1796}
  {\bibfield  {journal} {\bibinfo  {journal} {Nature Physics}\ }\textbf
  {\bibinfo {volume} {6}},\ \bibinfo {pages} {717} (\bibinfo {year}
  {2010})}\BibitemShut {NoStop}%
\bibitem [{\citenamefont {Gross}\ and\ \citenamefont
  {Sayama}(2009)}]{gross2009adaptive}%
  \BibitemOpen
  \bibfield  {author} {\bibinfo {author} {\bibfnamefont {T.}~\bibnamefont
  {Gross}}\ and\ \bibinfo {author} {\bibfnamefont {H.}~\bibnamefont {Sayama}},\
  }\href {http://books.google.de/books?id=P83skWv-c\_oC} {\emph {\bibinfo
  {title} {Adaptive Networks: Theory, Models and Applications}}},\
  Understanding Complex Systems\ (\bibinfo  {publisher} {Springer},\ \bibinfo
  {year} {2009})\BibitemShut {NoStop}%
\bibitem [{\citenamefont {Zhou}\ and\ \citenamefont {Kurths}(2006)}]{kurths06}%
  \BibitemOpen
  \bibfield  {author} {\bibinfo {author} {\bibfnamefont {C.}~\bibnamefont
  {Zhou}}\ and\ \bibinfo {author} {\bibfnamefont {J.}~\bibnamefont {Kurths}},\
  }\href@noop {} {\bibfield  {journal} {\bibinfo  {journal} {Phys. Rev. Lett.}\
  }\textbf {\bibinfo {volume} {96}} (\bibinfo {year} {2006})}\BibitemShut
  {NoStop}%
\bibitem [{\citenamefont {Ito}\ and\ \citenamefont
  {Kaneko}(2001)}]{itoKaneko01}%
  \BibitemOpen
  \bibfield  {author} {\bibinfo {author} {\bibfnamefont {J.}~\bibnamefont
  {Ito}}\ and\ \bibinfo {author} {\bibfnamefont {K.}~\bibnamefont {Kaneko}},\
  }\href {\doibase 10.1103/PhysRevLett.88.028701} {\bibfield  {journal}
  {\bibinfo  {journal} {Phys. Rev. Lett.}\ }\textbf {\bibinfo {volume} {88}},\
  \bibinfo {pages} {028701} (\bibinfo {year} {2001})}\BibitemShut {NoStop}%
\bibitem [{\citenamefont {Ravoori}\ \emph {et~al.}(2009)\citenamefont
  {Ravoori}, \citenamefont {Cohen}, \citenamefont {Setty}, \citenamefont
  {Sorrentino}, \citenamefont {Murphy}, \citenamefont {Ott},\ and\
  \citenamefont {Roy}}]{Ravoori09}%
  \BibitemOpen
  \bibfield  {author} {\bibinfo {author} {\bibfnamefont {B.}~\bibnamefont
  {Ravoori}}, \bibinfo {author} {\bibfnamefont {A.~B.}\ \bibnamefont {Cohen}},
  \bibinfo {author} {\bibfnamefont {A.~V.}\ \bibnamefont {Setty}}, \bibinfo
  {author} {\bibfnamefont {F.}~\bibnamefont {Sorrentino}}, \bibinfo {author}
  {\bibfnamefont {T.~E.}\ \bibnamefont {Murphy}}, \bibinfo {author}
  {\bibfnamefont {E.}~\bibnamefont {Ott}}, \ and\ \bibinfo {author}
  {\bibfnamefont {R.}~\bibnamefont {Roy}},\ }\href {\doibase
  10.1103/PhysRevE.80.056205} {\bibfield  {journal} {\bibinfo  {journal} {Phys.
  Rev. E}\ }\textbf {\bibinfo {volume} {80}},\ \bibinfo {pages} {056205}
  (\bibinfo {year} {2009})}\BibitemShut {NoStop}%
\bibitem [{\citenamefont {Zhigulin}\ \emph {et~al.}(2003)\citenamefont
  {Zhigulin}, \citenamefont {Rabinovich}, \citenamefont {Huerta},\ and\
  \citenamefont {Abarbanel}}]{Zhigulin}%
  \BibitemOpen
  \bibfield  {author} {\bibinfo {author} {\bibfnamefont {V.~P.}\ \bibnamefont
  {Zhigulin}}, \bibinfo {author} {\bibfnamefont {M.~I.}\ \bibnamefont
  {Rabinovich}}, \bibinfo {author} {\bibfnamefont {R.}~\bibnamefont {Huerta}},
  \ and\ \bibinfo {author} {\bibfnamefont {H.~D.~I.}\ \bibnamefont
  {Abarbanel}},\ }\href {\doibase 10.1103/PhysRevE.67.021901} {\bibfield
  {journal} {\bibinfo  {journal} {Phys. Rev. E}\ }\textbf {\bibinfo {volume}
  {67}},\ \bibinfo {pages} {021901} (\bibinfo {year} {2003})}\BibitemShut
  {NoStop}%
\bibitem [{\citenamefont {Sorrentino}\ and\ \citenamefont {Ott}(2009)}]{Ott09}%
  \BibitemOpen
  \bibfield  {author} {\bibinfo {author} {\bibfnamefont {F.}~\bibnamefont
  {Sorrentino}}\ and\ \bibinfo {author} {\bibfnamefont {E.}~\bibnamefont
  {Ott}},\ }\href {\doibase 10.1103/PhysRevE.79.016201} {\bibfield  {journal}
  {\bibinfo  {journal} {Phys. Rev. E}\ }\textbf {\bibinfo {volume} {79}},\
  \bibinfo {pages} {016201} (\bibinfo {year} {2009})}\BibitemShut {NoStop}%
\bibitem [{\citenamefont {Gorochowski}\ \emph {et~al.}(2010)\citenamefont
  {Gorochowski}, \citenamefont {di~Bernardo},\ and\ \citenamefont
  {Grierson}}]{Gorochowski10}%
  \BibitemOpen
  \bibfield  {author} {\bibinfo {author} {\bibfnamefont {T.~E.}\ \bibnamefont
  {Gorochowski}}, \bibinfo {author} {\bibfnamefont {M.}~\bibnamefont
  {di~Bernardo}}, \ and\ \bibinfo {author} {\bibfnamefont {C.~S.}\ \bibnamefont
  {Grierson}},\ }\href {\doibase 10.1103/PhysRevE.81.056212} {\bibfield
  {journal} {\bibinfo  {journal} {Phys. Rev. E}\ }\textbf {\bibinfo {volume}
  {81}},\ \bibinfo {pages} {056212} (\bibinfo {year} {2010})}\BibitemShut
  {NoStop}%
\bibitem [{\citenamefont {Aoki}\ and\ \citenamefont {Aoyagi}(2009)}]{aoki09}%
  \BibitemOpen
  \bibfield  {author} {\bibinfo {author} {\bibfnamefont {T.}~\bibnamefont
  {Aoki}}\ and\ \bibinfo {author} {\bibfnamefont {T.}~\bibnamefont {Aoyagi}},\
  }\href {\doibase 10.1103/PhysRevLett.102.034101} {\bibfield  {journal}
  {\bibinfo  {journal} {Phys. Rev. Lett.}\ }\textbf {\bibinfo {volume} {102}},\
  \bibinfo {pages} {034101} (\bibinfo {year} {2009})}\BibitemShut {NoStop}%
\bibitem [{\citenamefont {Seliger}\ \emph {et~al.}(2002)\citenamefont
  {Seliger}, \citenamefont {Young},\ and\ \citenamefont
  {Tsimring}}]{seliger02}%
  \BibitemOpen
  \bibfield  {author} {\bibinfo {author} {\bibfnamefont {P.}~\bibnamefont
  {Seliger}}, \bibinfo {author} {\bibfnamefont {S.~C.}\ \bibnamefont {Young}},
  \ and\ \bibinfo {author} {\bibfnamefont {L.~S.}\ \bibnamefont {Tsimring}},\
  }\href {\doibase 10.1103/PhysRevE.65.041906} {\bibfield  {journal} {\bibinfo
  {journal} {Phys. Rev. E}\ }\textbf {\bibinfo {volume} {65}},\ \bibinfo
  {pages} {041906} (\bibinfo {year} {2002})}\BibitemShut {NoStop}%
\bibitem [{\citenamefont {Bornholdt}\ and\ \citenamefont
  {Rohlf}(2000)}]{Bornholdt2000}%
  \BibitemOpen
  \bibfield  {author} {\bibinfo {author} {\bibfnamefont {S.}~\bibnamefont
  {Bornholdt}}\ and\ \bibinfo {author} {\bibfnamefont {T.}~\bibnamefont
  {Rohlf}},\ }\href {\doibase 10.1103/PhysRevLett.84.6114} {\bibfield
  {journal} {\bibinfo  {journal} {Phys. Rev. Lett.}\ }\textbf {\bibinfo
  {volume} {84}},\ \bibinfo {pages} {6114} (\bibinfo {year}
  {2000})}\BibitemShut {NoStop}%
\bibitem [{\citenamefont {Levina}\ \emph {et~al.}(2009)\citenamefont {Levina},
  \citenamefont {Herrmann},\ and\ \citenamefont {Geisel}}]{Levina09}%
  \BibitemOpen
  \bibfield  {author} {\bibinfo {author} {\bibfnamefont {A.}~\bibnamefont
  {Levina}}, \bibinfo {author} {\bibfnamefont {J.~M.}\ \bibnamefont
  {Herrmann}}, \ and\ \bibinfo {author} {\bibfnamefont {T.}~\bibnamefont
  {Geisel}},\ }\href {\doibase 10.1103/PhysRevLett.102.118110} {\bibfield
  {journal} {\bibinfo  {journal} {Phys. Rev. Lett.}\ }\textbf {\bibinfo
  {volume} {102}},\ \bibinfo {pages} {118110} (\bibinfo {year}
  {2009})}\BibitemShut {NoStop}%
\bibitem [{\citenamefont {Meisel}\ and\ \citenamefont
  {Gross}(2009)}]{Meisel09}%
  \BibitemOpen
  \bibfield  {author} {\bibinfo {author} {\bibfnamefont {C.}~\bibnamefont
  {Meisel}}\ and\ \bibinfo {author} {\bibfnamefont {T.}~\bibnamefont {Gross}},\
  }\href {\doibase 10.1103/PhysRevE.80.061917} {\bibfield  {journal} {\bibinfo
  {journal} {Phys. Rev. E}\ }\textbf {\bibinfo {volume} {80}},\ \bibinfo
  {pages} {061917} (\bibinfo {year} {2009})}\BibitemShut {NoStop}%
\bibitem [{\citenamefont {Magnasco}\ \emph {et~al.}(2009)\citenamefont
  {Magnasco}, \citenamefont {Piro},\ and\ \citenamefont
  {Cecchi}}]{Magnasco2009}%
  \BibitemOpen
  \bibfield  {author} {\bibinfo {author} {\bibfnamefont {M.~O.}\ \bibnamefont
  {Magnasco}}, \bibinfo {author} {\bibfnamefont {O.}~\bibnamefont {Piro}}, \
  and\ \bibinfo {author} {\bibfnamefont {G.~A.}\ \bibnamefont {Cecchi}},\
  }\href {\doibase 10.1103/PhysRevLett.102.258102} {\bibfield  {journal}
  {\bibinfo  {journal} {Phys. Rev. Lett.}\ }\textbf {\bibinfo {volume} {102}},\
  \bibinfo {pages} {258102} (\bibinfo {year} {2009})}\BibitemShut {NoStop}%
\bibitem [{\citenamefont {Guti\'errez}\ \emph {et~al.}(2011)\citenamefont
  {Guti\'errez}, \citenamefont {Amann}, \citenamefont {Assenza}, \citenamefont
  {G\'omez-Garde\~nes}, \citenamefont {Latora},\ and\ \citenamefont
  {Boccaletti}}]{Gutierrez2011}%
  \BibitemOpen
  \bibfield  {author} {\bibinfo {author} {\bibfnamefont {R.}~\bibnamefont
  {Guti\'errez}}, \bibinfo {author} {\bibfnamefont {A.}~\bibnamefont {Amann}},
  \bibinfo {author} {\bibfnamefont {S.}~\bibnamefont {Assenza}}, \bibinfo
  {author} {\bibfnamefont {J.}~\bibnamefont {G\'omez-Garde\~nes}}, \bibinfo
  {author} {\bibfnamefont {V.}~\bibnamefont {Latora}}, \ and\ \bibinfo {author}
  {\bibfnamefont {S.}~\bibnamefont {Boccaletti}},\ }\href {\doibase
  10.1103/PhysRevLett.107.234103} {\bibfield  {journal} {\bibinfo  {journal}
  {Phys. Rev. Lett.}\ }\textbf {\bibinfo {volume} {107}},\ \bibinfo {pages}
  {234103} (\bibinfo {year} {2011})}\BibitemShut {NoStop}%
\bibitem [{\citenamefont {Pecora}\ and\ \citenamefont
  {Carroll}(1990)}]{PecoraCarroll}%
  \BibitemOpen
  \bibfield  {author} {\bibinfo {author} {\bibfnamefont {L.~M.}\ \bibnamefont
  {Pecora}}\ and\ \bibinfo {author} {\bibfnamefont {T.~L.}\ \bibnamefont
  {Carroll}},\ }\href {\doibase 10.1103/PhysRevLett.64.821} {\bibfield
  {journal} {\bibinfo  {journal} {Phys. Rev. Lett.}\ }\textbf {\bibinfo
  {volume} {64}},\ \bibinfo {pages} {821} (\bibinfo {year} {1990})}\BibitemShut
  {NoStop}%
\bibitem [{\citenamefont {Pikovsky}\ \emph {et~al.}(2003)\citenamefont
  {Pikovsky}, \citenamefont {Rosenblum},\ and\ \citenamefont
  {Kurths}}]{synchronizationBook}%
  \BibitemOpen
  \bibfield  {author} {\bibinfo {author} {\bibfnamefont {A.}~\bibnamefont
  {Pikovsky}}, \bibinfo {author} {\bibfnamefont {M.}~\bibnamefont {Rosenblum}},
  \ and\ \bibinfo {author} {\bibfnamefont {J.}~\bibnamefont {Kurths}},\
  }\href@noop {} {\emph {\bibinfo {title} {Synchronization: A Universal Concept
  in Nonlinear Sciences}}}\ (\bibinfo  {publisher} {Cambridge University
  Press},\ \bibinfo {year} {2003})\BibitemShut {NoStop}%
\bibitem [{\citenamefont {Pecora}\ and\ \citenamefont
  {Carroll}(1998)}]{PecoraCaroll98}%
  \BibitemOpen
  \bibfield  {author} {\bibinfo {author} {\bibfnamefont {L.~M.}\ \bibnamefont
  {Pecora}}\ and\ \bibinfo {author} {\bibfnamefont {T.~L.}\ \bibnamefont
  {Carroll}},\ }\href {\doibase 10.1103/PhysRevLett.80.2109} {\bibfield
  {journal} {\bibinfo  {journal} {Phys. Rev. Lett.}\ }\textbf {\bibinfo
  {volume} {80}},\ \bibinfo {pages} {2109} (\bibinfo {year}
  {1998})}\BibitemShut {NoStop}%
\bibitem [{Note1()}]{Note1}%
  \BibitemOpen
  \bibinfo {note} {$f\left (x\right ) = r \protect \tmspace +\thinmuskip
  {.1667em} x \protect \tmspace +\thinmuskip {.1667em} \left (1-x\right )$ ,
  $r=4$}\BibitemShut {NoStop}%
\bibitem [{Note2()}]{Note2}%
  \BibitemOpen
  \bibinfo {note} {$f\left (x\right ) = \left (a \protect \tmspace +\thinmuskip
  {.1667em} x \right ) \protect \text {mod} \protect \tmspace +\thinmuskip
  {.1667em} 1$ , $a=\protect \frac {5}{3}$}\BibitemShut {NoStop}%
\bibitem [{Note3()}]{Note3}%
  \BibitemOpen
  \bibinfo {note} {Since $\epsilon \in \left [0,1\right ]$, synchronization
  thresholds $\epsilon _c > 1$ are purely hypothetical, although they can be
  calculated from the eigenvalues of the coupling matrix. They indicate that
  the correpsonding systems may not entirely synchronize.}\BibitemShut {Stop}%
\bibitem [{Note4()}]{Note4}%
  \BibitemOpen
  \bibinfo {note} {The adjacency matrix \protect \textbf {A} has entries
  $A^{ij} = 1$ if there is a link from unit $j$ to unit $i$. All other entries
  are zero.}\BibitemShut {Stop}%
\bibitem [{\citenamefont {Hong}\ \emph {et~al.}(2004)\citenamefont {Hong},
  \citenamefont {Kim}, \citenamefont {Choi},\ and\ \citenamefont
  {Park}}]{hongKimChoiPark}%
  \BibitemOpen
  \bibfield  {author} {\bibinfo {author} {\bibfnamefont {H.}~\bibnamefont
  {Hong}}, \bibinfo {author} {\bibfnamefont {B.~J.}\ \bibnamefont {Kim}},
  \bibinfo {author} {\bibfnamefont {M.~Y.}\ \bibnamefont {Choi}}, \ and\
  \bibinfo {author} {\bibfnamefont {H.}~\bibnamefont {Park}},\ }\href {\doibase
  10.1103/PhysRevE.69.067105} {\bibfield  {journal} {\bibinfo  {journal} {Phys.
  Rev. E}\ }\textbf {\bibinfo {volume} {69}},\ \bibinfo {pages} {067105}
  (\bibinfo {year} {2004})}\BibitemShut {NoStop}%
\end{thebibliography}

\end{document}